\documentclass[
reprint,
noeprint,
amsmath,amssymb,
aps,
pra,
floatfix,
superscriptaddress,
twocolumn,
10pt
]{revtex4-2}

\usepackage{color}
\usepackage{mathrsfs}
\usepackage{booktabs}
\usepackage{amsthm}
\usepackage{hyperref}
\hypersetup{colorlinks=true, citecolor=blue, urlcolor=blue, linkcolor=blue}

\usepackage{mathtools} 
\usepackage{graphicx}
\usepackage{dcolumn}
\usepackage{bm}
\usepackage{dsfont}
\usepackage{etoolbox}
\apptocmd{\sloppy}{\hbadness 10000\relax}{}{}
\usepackage{braket}
\newcommand{\be}{\begin{equation}}
\newcommand{\ee}{\end{equation}}
\begin{document}

\preprint{APS/123-QED}

\title{Adiabatic modulation of driving protocols in periodically driven quantum systems}

\author{Ashwin \surname{Murali}}
 \email{am957phys@gmail.com}
 \affiliation{Department of Chemistry, Birla Institute of Technology and Science, Pilani.}
 
\author{Tapomoy \surname {Guha Sarkar}}%
 \email{tapomoy@pilani.bits-pilani.ac.in}
\affiliation{
 Department of Physics, Birla Institute of Technology and Science, Pilani.
 }

\author{Jayendra N. Bandyopadhyay}
    \email{jayendra@pilani.bits-pilani.ac.in}
\affiliation{
Department of Physics, Birla Institute of Technology and Science, Pilani.
}
\providecommand{\abs}[1]{\lvert#1\rvert}
\renewcommand{\d}{\mathrm{d}}
\renewcommand{\i}{\mathrm{i}}

\begin{abstract}
 We consider a periodically driven system where the high-frequency driving protocol consists of a sequence of potentials switched on and off at different instants within a period. We explore the possibility of introducing an adiabatic modulation of the driving protocol by considering a slow evolution of the instants when the sequence of potentials is switched on/off. We examine how this influences the long-term dynamics of periodically driven quantum systems. By assuming that the slow and fast timescales in the problem can be decoupled, we derive the stroboscopic (effective) Hamiltonian for a four-step driving sequence up to the first order in perturbation theory. We then apply this approach to a spin-$1/2$ system, where the adiabatic modulation of the driving protocol is chosen to produce an evolving emergent magnetic field that interacts with the spin. We study the emergence of \emph{diabolical points} and \emph{diabolical loci} in the parameter space of the effective Hamiltonian. Further, we study the topological properties of the maps of the adiabatic paths in the parameter space to the eigenspace of the effective Hamiltonian. In effect, we obtain a technique for tuning the topological properties of the eigenstates by selecting different adiabatic evolutions of the driving protocol, characterized by distinct paths in parameter space. This technique can be applied to any periodic driving protocol to achieve desirable topological effects.
\end{abstract}

\maketitle


\section{Introduction}
Floquet engineering \cite{weitenberg2021, rudner2020fe, Oka2019, eckardt2017} of quantum systems has emerged as a powerful tool to synthesize exotic phases of matter \cite{mori2023, rudner2020, else2016, vonkeyserlingk2016, cayssol2013, Rechtsman2013}, and has diverse applications in quantum simulations using ultracold atoms \cite{wintersperger2020, reichl2014, jiang2011} and realizations of artificial gauge fields \cite{schweizer2019, bukov2016, hauke2012, creffield2011}. The essential strategy behind Floquet engineering relies on the fact that the long-term evolution of a periodically driven quantum system is governed by a static effective Hamiltonian, barring an initial and final micro-motion. The approach involves a choice of an undriven system and then suitably tailoring the driving scheme to design effective Hamiltonians with desired properties \cite{rahav2003effective, goldman2014periodically, eckardt2015}.

A class of systems involving the coupling of a spin with a magnetic field is of great interest as they demonstrate non-trivial topological properties \cite{arakawa2021, bukov2015, galitski2013, hemmerich2010, sorensen2005}. The eigenstates of such systems generally acquire a geometrical phase \cite{pancharatnam1956, berry1984, aharanov1987, samuel1988, mukunda1993}  when slowly transported around a closed path by varying the parameters $\vec{R}$ in its Hamiltonian $\widehat{\mathcal{H}}(\vec{R})$. Further, degeneracies in such systems with at least two parameters also exhibit \emph{diabolical points} with double-cone-like structures at such points in the band diagram \cite{hsin2020, berry1984diabolical, shanley1990}. In traditional studies of adiabatic evolution \cite{kolodrubetz2018, zhou2016}, the parameter space consists of the constants representing the strengths of different operators appearing in the Hamiltonian. It is usually not possible to increase the number of parameters without introducing new operators, thereby limiting the complexity of the dynamics. 
We suggest a novel method of synthesizing a Hamiltonian where the number of parameters can be increased without requiring any new basic operators as the building blocks for the Hamiltonian.

We achieve this in two steps. Firstly, we consider a suitably chosen static Hamiltonian and design a high-frequency periodic driving protocol to engineer the desired properties in an effective Hamiltonian \cite{goldman2014periodically, eckardt2015, rahav2003effective}. 
Secondly, if the driving protocol is envisaged to be a sequence of operators switched on/off over time slices within a full period, we consider the switch on/off times of the operators as the relevant parameters. Since the slicing of the time period can be done in innumerable ways, and the switch on/off time appears in the effective Hamiltonian, we can engineer a Hamiltonian in an enhanced parameter space corresponding to the large number of slices of the time period. These parameters of the effective Hamiltonian are then adiabatically evolved for interesting geometric consequences. 

We, thus, have two time scales in our problem: the fast scale which generates the effective Hamiltonian, and the slow scale, over which the protocol itself evolves. 
We note that the strengths of the operators in the driving protocol also appear in the effective Hamiltonian. However, we do not consider these as our adiabatically evolving parameters as is done in \cite{nathan2021, kolodrubetz2018, zhou2016}. Thus, the strengths of the different potentials are held constant, whereas their switch on/off times are our adiabatically evolving parameters.
Multi-frequency/quasi-periodic driving of quantum systems is usually studied using the notion of Floquet lattice \cite{park2023, long2022}. However, the time scales involved in our analysis are so widely separated that we have decoupled the two time evolutions in our analysis. We demonstrate our method using the simplest spin system.

In this paper, we first consider a high-frequency driving of a spin-$1/2$ system with a protocol consisting of a four-step sequence of potentials switched on and off at different instants within a period. We further invoke a much slower time scale for an adiabatic modulation of the driving protocol itself through a slow variation of the instants when the sequence of potentials is switched on/off. The choice of operators in the driving protocol produces an evolving emergent magnetic field that interacts with the spin. We then look at adiabatic evolution in the space of the parameters of the driving protocol, which effectively maps to the evolution of the synthesized magnetic field.  We study the emergence of \emph{diabolical points} and topological properties of adiabatic paths. The geometric phase picked up by the eigenstates during the adiabatic evolution and its variation for different forms of adiabatic protocols is subsequently studied. In effect, we obtain a technique to tune the winding of the eigenstates and the accumulated geometric phase by choosing various adiabatic evolution of the driving protocol characterized by diverse paths in the parameter space.

\section{Formalism}
A periodically driven quantum system is characterized by a time-dependent Hamiltonian $\widehat{\mathcal{H}} (t) = \widehat{\mathcal{H}}(t + T)$,
where $T$ is the driving period.
We consider systems where the full Hamiltonian can be written as a time-independent 
base Hamiltonian $\widehat{\mathcal{H}}_0$ driven by a time-dependent
external potential $\widehat{V}(t)$ which acts as a time-periodic perturbation to the
system. The total Hamiltonian is,  thus, written as 
\begin{equation}
    \widehat{\mathcal{H}}(t) = \widehat{\mathcal{H}}_0 + \widehat{V}(t), ~\text{with} \ \widehat{V}(t+T) = \widehat{V}(t).
\end{equation}
 The dynamics of  such a system in an initial state $|\psi(t_i)\rangle$,
 can described in the period $t_i \rightarrow t_f$ 
using the propagator $\widehat{\mathcal{U}}(t_i, t_f)$ as $|\psi(t_f)\rangle = \widehat{\mathcal{U}}(t_i, t_f)|\psi(t_i)\rangle$ where the unitary operator $\widehat{\mathcal{U}} $ is given by  \cite{eckardt2015, goldman2014periodically, rahav2003effective}
\begin{equation}
\widehat{\mathcal{U}}(t_i, t_f)
= e^{-\i\widehat{\mathcal{K}}(t_f)}e^{-\i\widehat{\mathcal{H}}_{\text{eff}}(t_f - t_i)}e^{\i\widehat{\mathcal{K}}(t_i)}.
\end{equation}
Here, $\widehat{\mathcal{K}}(t)$ is a time-periodic operator with period $T$ known
as the micro-motion kick operator, and $\widehat{\mathcal{H}}_{\text{eff}}$ is a time-independent
effective Hamiltonian. The expressions for the kick operator and the effective
Hamiltonian can be obtained using a perturbative expansion \cite{goldman2014periodically, rahav2003effective}
in the limit of high driving frequency $\omega = \frac{2\pi}{T} >> \omega_0$, where $\hbar\omega_0$ is the typical
energy scale of the base Hamiltonian $\widehat{\mathcal{H}}_0$. 

For high-frequency driving, the quantity $1/\omega$ can be treated as a perturbation parameter and 
the operators $\widehat{\mathcal{H}}_{\text{eff}}$ and
$\widehat{\mathcal{K}} (t)$ may be expanded as \cite{rahav2003effective, goldman2014periodically}

  \begin{align}
  &\widehat{\mathcal{H}}_{\text{eff}} = \widehat{\mathcal{H}}_0 + \sum_{m=1}^{\infty} \frac{\widehat{\mathcal{H}}_m}{\omega^m} \text{, and} \quad \widehat{\mathcal{K}} (t) = \sum_{m=1}^{\infty} \frac{\widehat{\mathcal{K}}_{m}(t)}{\omega^m}.
\end{align}

The effective Hamiltonian gives a stroboscopic description of the dynamics and can be engineered to produce desirable properties \cite{Oka2019, rudner2020fe}.
The above analysis can be applied to a specific class of driving protocols in which the potential $\widehat{V}(t)$ is a piecewise constant function of time with $N$ pieces (or steps)
in each period $T$. 
In this protocol
$\Pi :\{\widehat{V}_1, \widehat{V}_2, \ldots, \widehat{V}_N\}$, 
 the potentials $\widehat{V}_r$  are switched  on for a
duration of $\frac{T}{N}$ and then  switched off sequentially in each period $T$. Such equi-timed sequences of potentials are studied extensively in \cite{goldman2014periodically}.

Instead of considering equi-timed steps we introduce two parameters $(\alpha, \beta)$  $\in [0, 1]$ 
in a four-step protocol $\Pi(\alpha, \beta)  :\{\widehat{V}_1, \widehat{V}_2, \widehat{V}_3, \widehat{V}_4\}$ defined as:
\begin{align}\label{eq:protocol}
&\widehat{V}(t) = \begin{cases*}
  \widehat{V}_1 &  $0 \leq t \leq \frac{\alpha T}{2}$  \\
  \widehat{V}_2 & $\frac{\alpha T}{2} \leq t \leq \frac{T}{2}$ \\
  \widehat{V}_3 & $\frac{T}{2} \leq t \leq \frac{T(1 + \beta)}{2}$ \\
  \widehat{V}_4 & $\frac{T(1 + \beta)}{2} \leq t \leq T$
\end{cases*}.
\end{align}
These parameters fix the switch on/off time for the potentials within a full period and hence characterize the driving protocol along with the potentials themselves. 
To simplify our calculation, we impose a condition $\sum_{r = 1}^{N} \widehat{V}_r = 0$.
We shall use the protocol in Eq.~(\ref{eq:protocol}) to engineer effective Hamiltonians for high-frequency driving of quantum systems. 
\begin{figure}[htbp!] 
  \centering
\includegraphics[width=\linewidth]{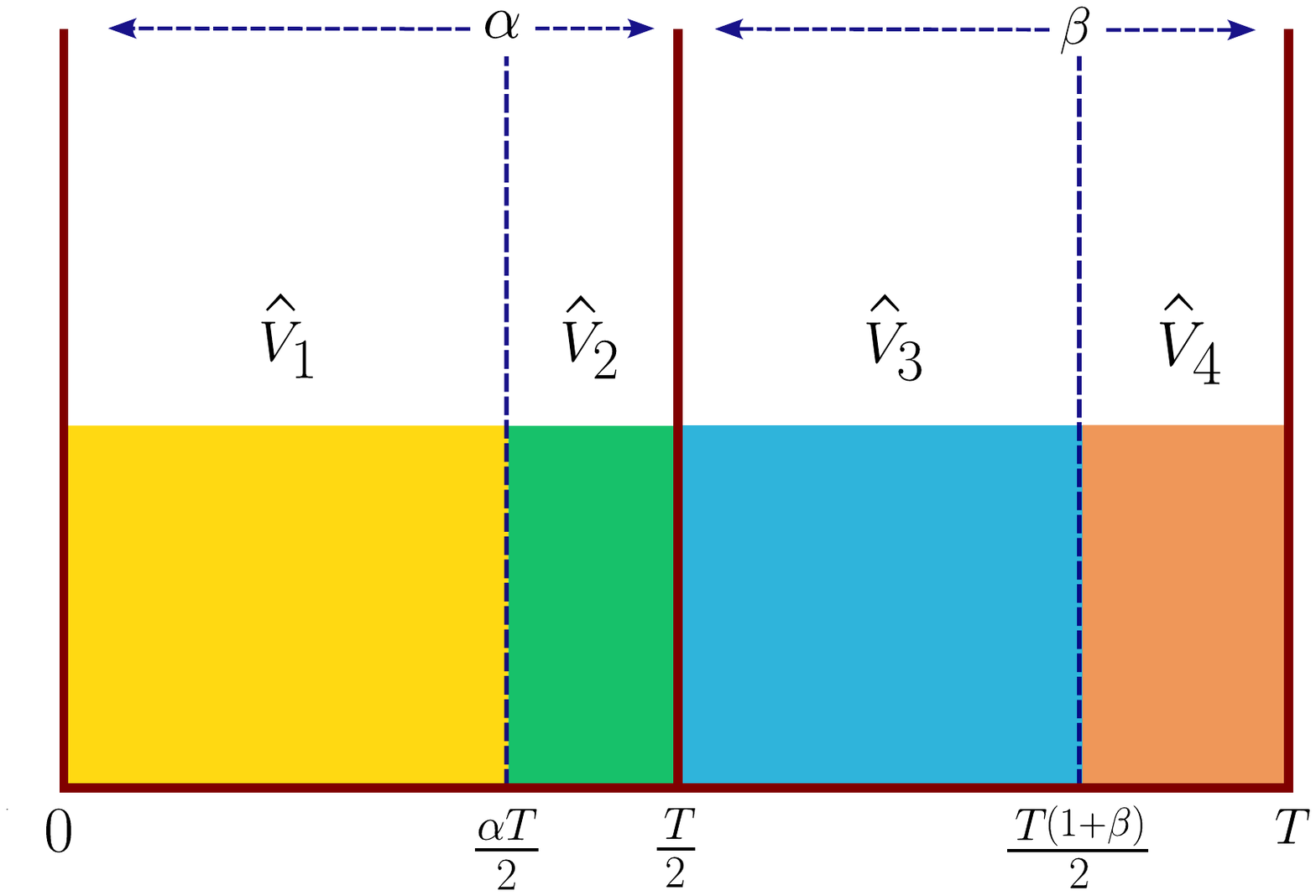}
  \caption{The protocol $\Pi(\alpha, \beta)$ parameterized by two parameters $\alpha$ and $\beta$.}
\end{figure}

We follow the formalism in Appendix C of \cite{goldman2014periodically} where the perturbative terms $\widehat{\mathcal{H}}_m$ of the effective Hamiltonian $\widehat{\mathcal{H}}_{\text{eff}} $ are computed using the Fourier components $V^{(j)}$ of the driving potential  defined through 
\be\label{eq:fourier}
V^{(j)} = \frac{1}{T}\int_{0}^{T}\widehat{V}(t)e^{- \i j\omega t}dt , 
\ee
These Fourier components depend on the parameters  $(\alpha, \beta)$ through their appearance in the integration limits for
$\widehat{V}(t)$ in  Eq.~(\ref{eq:protocol}).  Consequently, this shall imprint the parameters onto $\widehat{\mathcal{H}}_m$. 

The modulated effective Hamiltonian up to the first order is given by  
\begin{equation}
\label{eq:master}
\widehat{\mathcal{H}}_{\text{eff}} = \widehat{\mathcal{H}}_0 + \frac{\i\pi}{8\omega}
\sum_{\substack{r, s = 1\\ s > r}}^{4} \mathscr{P}_{rs}\left(\alpha, \beta\right)\left[\widehat{V}_r, \widehat{V}_s\right]
+ \mathcal{O} \left(\omega^{-2} \right), \end{equation}
where the six bi-variate polynomials $\mathscr{P}_{rs}$ are given by:
\begin{align}
    &\mathscr{P}_{12}(\alpha, \beta) = \alpha(1- \alpha),~ \mathscr{P}_{13}(\alpha, \beta) = \alpha\beta(\alpha - \beta),\nonumber\\ 
    &\mathscr{P}_{14}(\alpha, \beta) = \alpha(1 - \beta)(\alpha - \beta - 1), \nonumber\\
    & \mathscr{P}_{23}(\alpha, \beta) = \beta(1-\alpha)(\alpha - \beta + 1),\nonumber\\ 
    &\mathscr{P}_{24}(\alpha, \beta) = (\alpha -1)(\beta -1)(\alpha - \beta), \nonumber \\
    &\mathscr{P}_{34}(\alpha, \beta) = \beta(1 - \beta).
\end{align}
We note that for $\left(\alpha, \beta\right) = \left(\frac{1}{2}, \frac{1}{2}\right)$ this reduces to the equi-timed four-step protocol \cite{goldman2014periodically}. For four-step protocols, where  $\left(\alpha, \beta\right) \neq \left(\frac{1}{2}, \frac{1}{2}\right)$,
additional commutators $[\widehat{V}_r, \widehat{V}_s]$ emerge as opposed to the case when $\left(\alpha, \beta\right) = \left(\frac{1}{2}, \frac{1}{2}\right)$ where only commutators of  the form $[\widehat{V}_r, \widehat{V}_{(r~{\text {mod}}~ 4 )+ 1}]$ appear in the effective Hamiltonian. 

Similarly, the  micro-motion kick operator up to the first order may be obtained as 
\begin{equation}
    \widehat{\mathcal{K}}(0) = -\frac{\pi}{4\omega}\sum_{r=1}^{4}\mathscr{Q}_r\left(\alpha, \beta\right)\widehat{V}_r + \mathcal{O} \left(\omega^{-2} \right),
\end{equation}
where the four bi-variate polynomials $\mathscr{Q}_{r}\left(\alpha, \beta\right)$ are given by
\begin{align}
    &\mathscr{Q}_{1}\left(\alpha, \beta\right) = \alpha\left(2 - \alpha\right),~\mathscr{Q}_{2}\left(\alpha, \beta\right) = \left(\alpha - 1\right)^2 , \nonumber\\
    &\mathscr{Q}_{3}\left(\alpha, \beta\right) = -\beta^2,~\mathscr{Q}_{4}\left(\alpha, \beta\right) = \left(\beta^2 -1\right).
\end{align}
We find that the polynomials $\mathscr{Q}_r\left(\alpha, \beta\right)$s  are either functions of $\alpha$ or $\beta$. This is expected since the first-order correction here is a sum over the potentials and not over products of potentials like it is for the effective Hamiltonian. Thus, the factors $\alpha$ and $\beta$ do not appear as a product.

The effective Hamiltonian $\widehat{\mathcal{H}}_{\text{eff}}$ is obtained by averaging over the high-frequency variation of the potential. 
The fast driving protocol can be modified by changing $(\alpha, \beta)$ through changes in the values of polynomials $\mathscr{P}_{rs}(\alpha, \beta)$.
We therefore consider another time scale $\tau  >> \omega^{-1}$ over which $(\alpha, \beta)$ evolve.
This results in an adiabatic evolution of $\widehat{\mathcal{H}}_{\text{eff}} ( \vec R(\tau))$, where $\vec{R}(\tau) = ( \alpha(\tau), \beta(\tau) )   \in [0,1] \times [0,1]$ evolves smoothly and slowly in the parameter space.
We use $\tau$ to indicate that we looking at adiabatic variations different from the fast stroboscopic time scale $t$ varying over $T$.
This is akin to Berry's adiabatic evolution \cite{berry1984} of a Hamiltonian in the parameter space, except that our parameters are now the characteristics of a fast-driving protocol.
The adiabatic evolution $\vec{R}(\tau) = ( \alpha(\tau), \beta(\tau) ) $  is a path in the 2D parameter space $ [0,1] \times [0,1] \subset {\mathbb R}^2$. We parameterize this path by a continuous
vector-valued map 
\begin{align}
\vec{R} : &[a, b] \rightarrow [0, 1]\times[0, 1]\nonumber\\
&\tau \mapsto \left(\alpha(\tau), \beta(\tau)\right).
\end{align}
We are interested in the possibility that the eigenstates of $\widehat{\mathcal{H}}_{\text{eff}} ( \vec R(\tau)) $ pick up a nontrivial geometric phase under such adiabatic evolution.
\subsection{Driven spin $1/2$ system}
\begin{figure*}
  \centering
  \includegraphics[width=12.5cm, height=5.5cm]{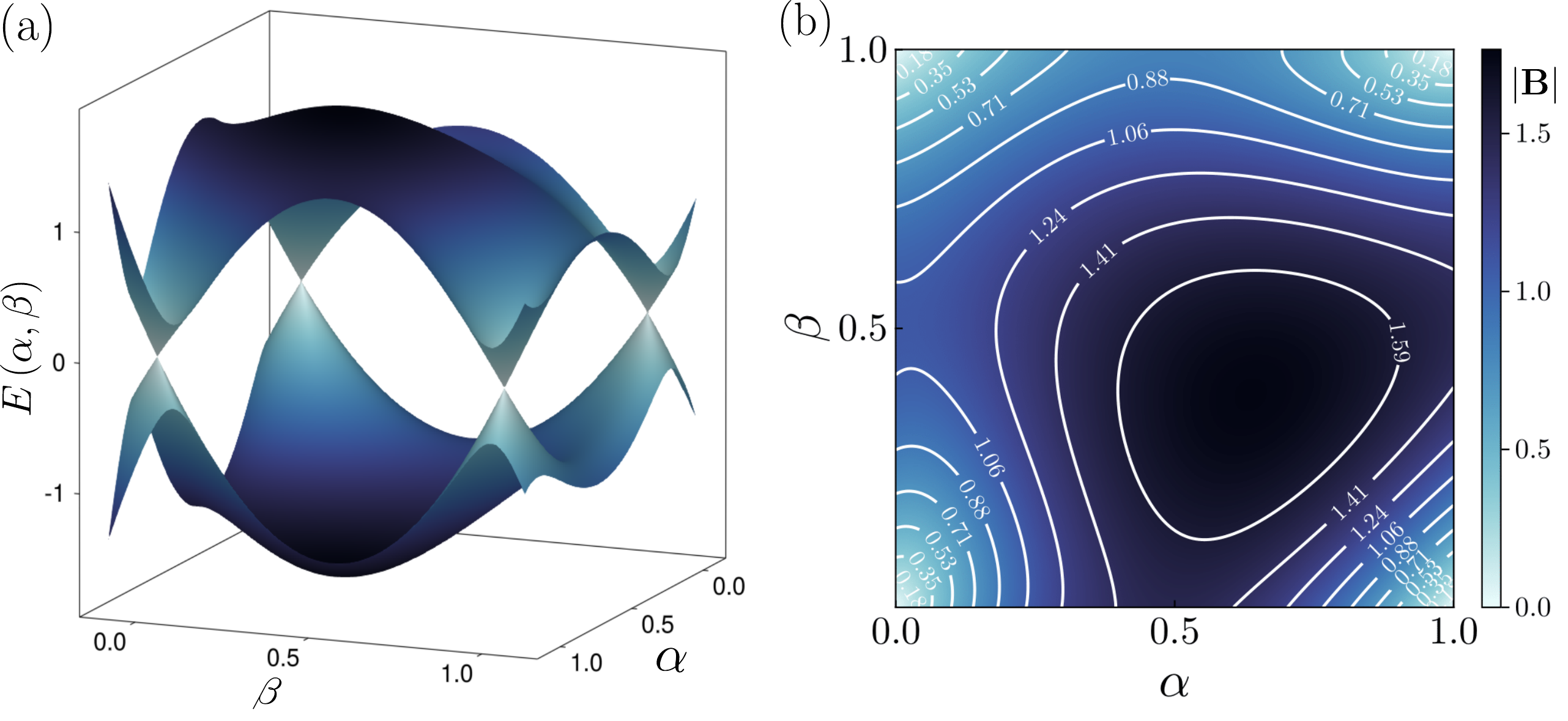}
        \caption{ (a) The band structure of the effective Hamiltonian for the protocol with $\left(c_1, c_2, c_3, c_4\right) = (1,1,1,1)$. (b) The magnitude of the synthetic magnetic field over the domain of parameters
  $\alpha$ and $\beta$. }
  \label{fig:band}
  \end{figure*}
\begin{figure*}
  \centering
  \includegraphics[width=12.5cm, height=5.5cm]{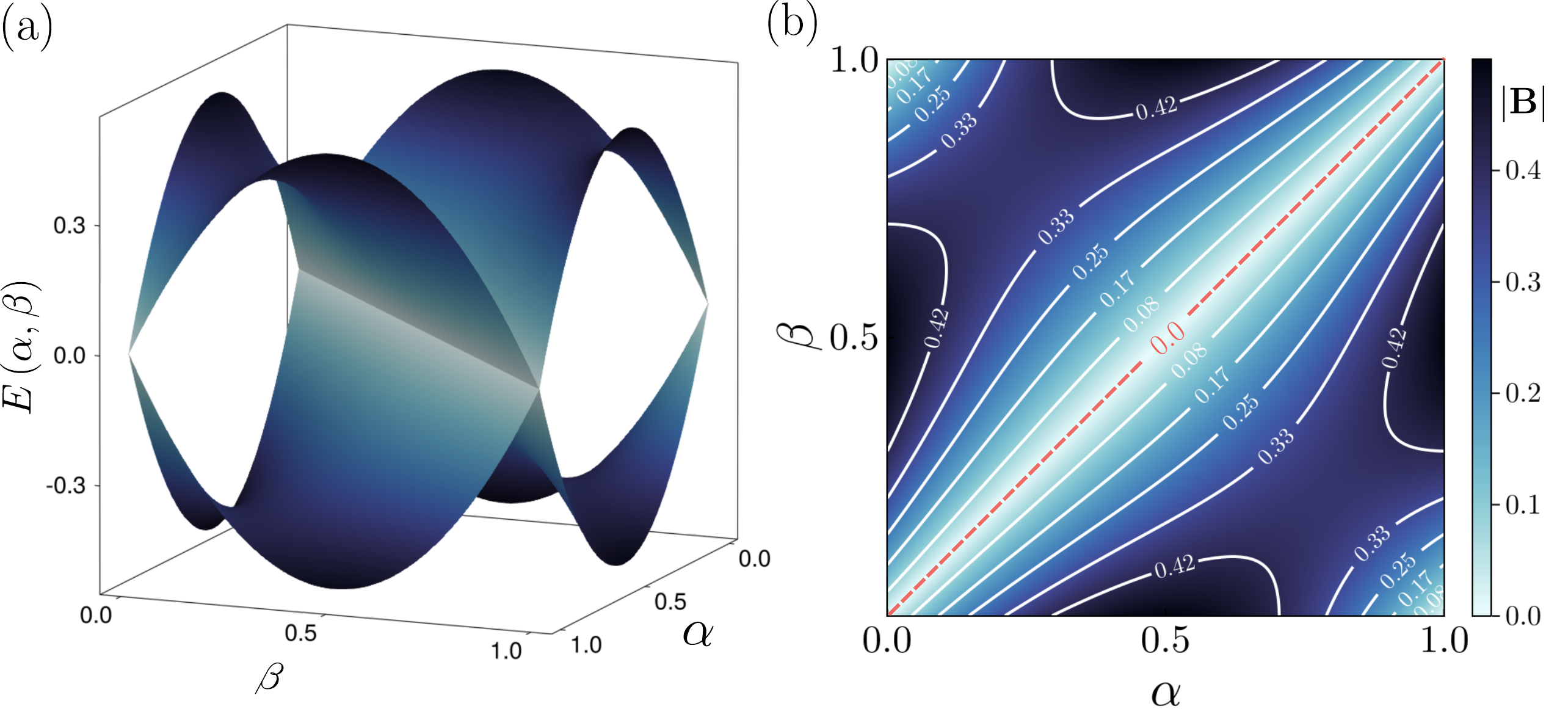}
        \caption{ (a) The band structure of the effective Hamiltonian for the protocol with $\left(c_1, c_2, c_3, c_4\right) = (0,1,0,1)$. (b) The magnitude of the synthetic magnetic field over the domain of parameters
  $\alpha$ and $\beta$. The dashed line (red) along the diagonal corresponds to the loci of gapless points.}
  \label{fig:band2}
  \end{figure*}
Adiabatic evolution of quantum Hamiltonian systems \cite{born1928, kato1950} in the parameter space is comprehensively investigated for undriven Hamiltonians, where the adiabatic parameters are restricted to the parameters of the stationary Hamiltonian 
in the manner of Berry \cite{berry1984}. 
We are interested in driven systems and aim to use the parameters of the driving protocol to induce adiabatic evolution of an effective Hamiltonian. 
One of the simplest systems demonstrating anholonomy is the two-level spin-$1/2$ system, subjected to an external magnetic field, for which the ground state picks up a non-zero Berry phase on adiabatically varying the direction
of the magnetic field. 
This motivates us to engineer a driving protocol that
would give us an effective Hamiltonian like the anholonomy Hamiltonian
of the two-state system. That is, we desire that the engineered Hamiltonian is of the form $\mathbf{S}\cdot \mathbf{B}$, where $\mathbf{B}$ is an effective magnetic field and $\mathbf{S} = \left(S_x, S_y, S_z\right)$ is the spin angular momentum operator. 

We choose our undriven system to be a two-level system of a spin-$1/2 (s=1/2)$, particle, such that $\widehat{\mathcal{H}}_0 = 
\frac{3}{4}\frac{\mathds{1}_{\text{2x2}}}{2I}$
in the $S_z$ basis. The dimensionful number $I$ sets the scale of energy for the undriven system. We drive this system with a protocol $\Pi\left(\alpha, \beta, c_i\right)$ with the potentials given by:
\begin{align}
&\widehat{V}_1 = c_1 S_x - c_2 S_y \ , \widehat{V}_2 = c_2 S_y + c_3 S_z ,\nonumber \\
&\widehat{V}_3 = c_4 S_z - c_1 S_x \ , \widehat{V}_4 =  -(c_3 + c_4) S_z , 
\end{align}
where $c_1, c_2, c_3, c_4$ are suitable constants. It follows from Eq.~(\ref{eq:master}) that the first perturbative correction to the effective Hamiltonian will just be a linear combination of the spin operators. We may interpret the coefficients of the spin operators as the components of a synthetic magnetic field $\mathbf{B}$. For such a protocol, the effective Hamiltonian and micro-motion kick operator takes the desired form 
\begin{align}
    \widehat{\mathcal{H}}_{\text{eff}} &= \frac{S^2}{2I} -\frac{\pi}{8\omega}
    \mathbf{S}\cdot \mathbf{B} \quad \text{and} \quad  \widehat{\mathcal{K}}\left(0\right) = -\frac{\pi}{4\omega} \mathbf{S}\cdot \mathbf{B'},
\end{align}
where $\mathbf{B}$ and $\mathbf{B^{'}}$ are given by
\begin{widetext}
  \begin{equation}
  \mathbf{B} = 
  \begin{bmatrix}
    c_2 c_3\left(-\mathscr{P}_{12} + \mathscr{P}_{14} - \mathscr{P}_{24} \right) 
+ c_2 c_4\left(-\mathscr{P}_{13} + \mathscr{P}_{23} - \mathscr{P}_{24} + \mathscr{P}_{14}\right)\\
    c_1 c_3\left(-\mathscr{P}_{12} + \mathscr{P}_{14} - \mathscr{P}_{34} - \mathscr{P}_{23}\right) + c_1 c_4\left(-\mathscr{P}_{13} - \mathscr{P}_{34} + \mathscr{P}_{14}\right)\\
    c_1 c_2\left(\mathscr{P}_{12} - \mathscr{P}_{13} + \mathscr{P}_{23}\right)\\
  \end{bmatrix},
~ \mathbf{B'}= 
\begin{bmatrix}
  c_1\left(\mathscr{Q}_1 - \mathscr{Q}_3\right)\\
  c_2\left(\mathscr{Q}_2 - \mathscr{Q}_1\right) \\
  c_3\mathscr{Q}_2 + c_4\mathscr{Q}_3 - \left(c_3+c_4\right)\mathscr{Q}_4\\
  \end{bmatrix}
  \end{equation}
  \end{widetext}
We note that with our choice of $\widehat{\mathcal{H}}_0 $
the perturbative corrections to the energy at and above $\mathcal{O} \left(\omega^{-2} \right) $ vanish.

\section{Results}
\label{sec:domain}
\subsection{The Energy Spectrum}
The eigenvalues of the effective Hamiltonian $\widehat{\cal{H}}_\text{eff}$ as a function of the parameters $(\alpha, \beta)$ are given by 
$E\left(\alpha, \beta\right) = \frac{3}{8I} \pm \frac{\pi}{8\omega} \abs{\mathbf{B}(\alpha, \beta)}$. Upto an overall shift of energy by $\frac{3}{8I}$ and scaling by $\frac{\pi}{8\omega}$ the spectrum is determined by $\pm {\abs{\mathbf{B(\alpha, \beta)}}}$. We use units $\hbar = c = 1$ in our analysis, such that $E\left(\alpha, \beta\right)$ and $\abs{\mathbf{B\left(\alpha, \beta\right)}}$ are dimensionless quantities with all energies in units of $I^{-1}$. A choice of parameters $(\alpha, \beta)$, thus allows us to tune the spectrum by causing level repulsion or reduction of the energy gap. 
The gapless degenerate points are particularly important since they are topologically protected.

It is well known that for a spin-1/2 Hamiltonian in a 2-dimensional parameter space, the degeneracies of the ground state form a sub-manifold of the parameter space. These correspond to gapless \emph{diabolical points} and/or \emph{loci} (joining the \emph{diabolical points}) as studied in \cite{hsin2020, berry1984diabolical}. For $(\alpha, \beta) \in \{ (0, 0), (0, 1), (1, 0), (1, 1) \} $, all the polynomials $\mathscr{P}_{rs}(\alpha, \beta) = 0$, whereby the magnetic field vanishes. These points at the four corners of the parameter space where the bands touch are thus the \emph{diabolical points} for this system and remain preserved irrespective of our choice of constants $\left(c_1, c_2, c_3, c_4\right)$ in the protocol $\Pi$.
We note that at these points, our four-step protocol behaves like a two-step protocol since two of the pulse sequences overlap with each other, resulting in only two of the four potentials being switched on/off during the driving period. Further,  changing the constants, $\left(c_1, c_2, c_3, c_4\right)$ can not result in the creation of new \emph{diabolical points}. However, singular gapless lines connecting these points may emerge for a suitable choice of these constants. 

We study two scenarios where: (i) there are only the four isolated \emph{diabolical points} (ii) there are additional diabolical curves connecting these points. One way of realizing the first scenario is to choose the constants to be $\left(c_1, c_2, c_3, c_4\right) =(1,1,1,1)$. While this choice is not special by any means, it encapsulates some generic aspects of the scenario.
For example, when the constants are all positive, the qualitative behavior is not different from this.

The Floquet engineered magnetic field $\mathbf{B}(\alpha, \beta)$ for $\left(c_1, c_2, c_3, c_4\right) =(1,1,1,1)$ is given by 
\begin{equation}
  \mathbf{B} = \begin{bmatrix}
    -6\alpha^2\beta + 5\alpha^2 + 6\alpha\beta^2 -5\alpha -3\beta^2 + 3\beta \\
    -2\alpha^2\beta + 3\alpha^2 + 2\alpha\beta^2 -3\alpha + 3\beta^2 - 3\beta \\
    -2\alpha^2\beta - \alpha^2 + 2\alpha\beta^2 +\alpha - \beta^2 +\beta \\
  \end{bmatrix}
  \end{equation}
Figure \ref{fig:band}(a) shows the two-band structure of the energy spectrum as a function of the parameters. Figure \ref{fig:band}(b) shows the magnitude of the possible magnetic fields. The magnitude of the magnetic field attains a maximum near the center for  $(\alpha, \beta)=(0.63, 0.38)$. The band structure in the vicinity of the four \emph{diabolical points} exhibits an asymmetrical conical structure, as seen in Fig.~\ref{fig:band}(a).

It is, in general, possible to conceive of protocols where we choose the values of $\left(c_1, c_2, c_3, c_4\right)$
to generate gapless loci along with the 4 \emph{diabolical points} in the parameter space.
For example, a choice of $\left(c_1, c_2, c_3, c_4\right) = (0,1,0,1)$ achieves this.
Figure \ref{fig:band2} shows the band diagram for this case where the engineered magnetic field is of the form  $\mathbf{B} = \left ( -2\alpha + 2\alpha^2 + 2\beta - 4 \alpha^2\beta - 2\beta^2 + 4 \alpha \beta^2 \right ) \hat{x}$.
This magnetic field vanishes not only at the four corners but also along the line $\alpha = \beta$.
Figure \ref{fig:band2}(b) shows the magnitude of the magnetic field where we observe the gapless loci connecting the \emph{diabolical points} $(0,0)$ and $(1,1)$ along the leading diagonal across the parameter space.

\subsection{Adiabatic Paths}
\begin{figure}[htbp!]
  \includegraphics[scale=0.24]{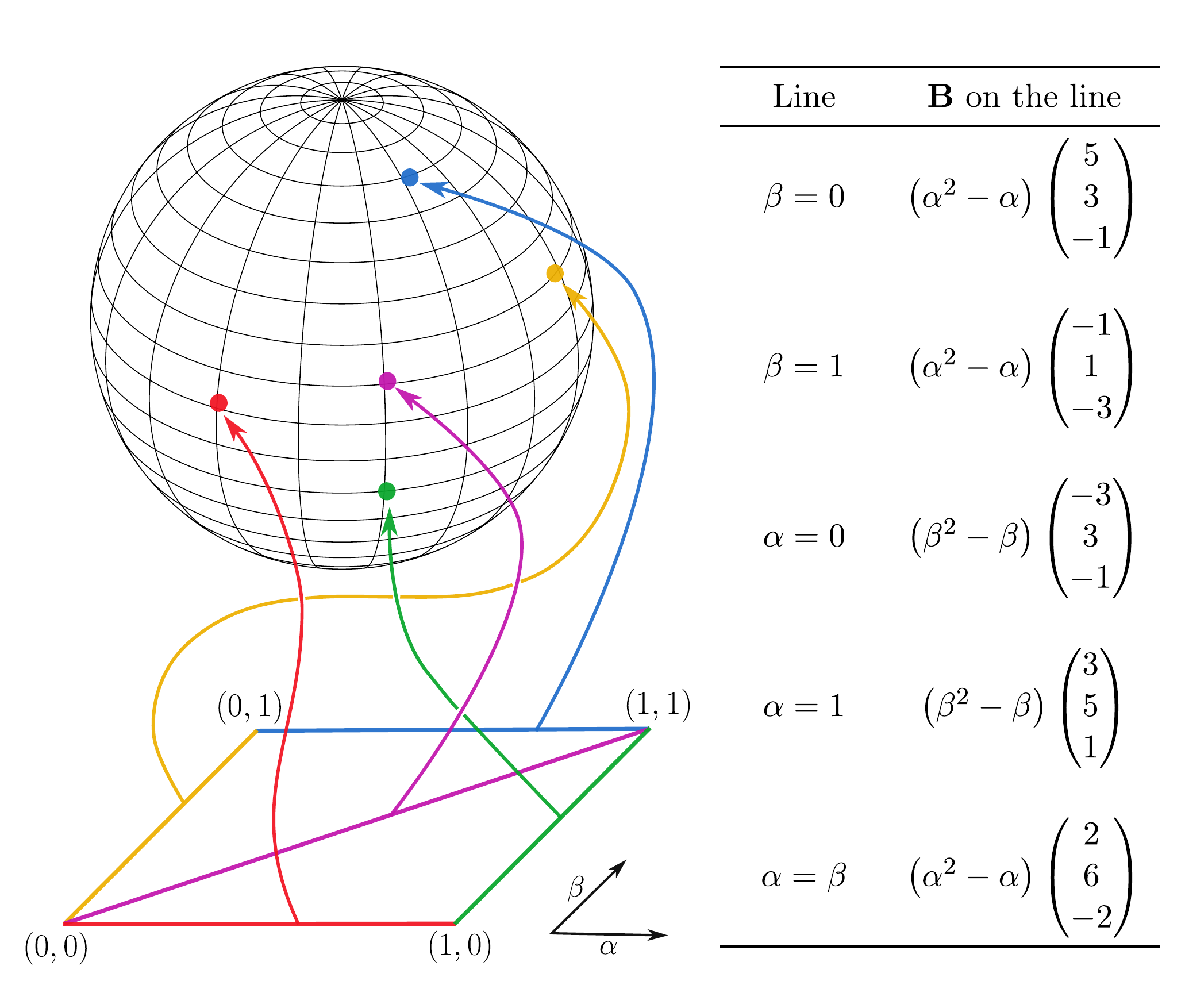}
  \caption{  A schematic diagram depicting the mapping of the boundary of the parameter space to the unit sphere where $\hat B = (\theta, \phi)$, for $\left(c_1, c_2, c_3, c_4\right) = \left(1, 1, 1, 1\right)$. The table lists the magnetic field on each invariant line on the domain.}
     \label{fig:sphere}
\end{figure}
The ground state $| \chi \rangle $ of the effective Hamiltonian can be represented by a point on $(\theta, \phi)$ the unit Bloch sphere, which corresponds to the direction $\hat{B}$  of the engineered magnetic field. 
Thus, an adiabatic evolution of the parameters  $\vec{R}\left(\tau \right)$ in the unit-square maps to a path $\vec{r}\left(\tau\right)$ traced by the unit vector $\hat{n} = {\mathbf{B}}/ {|\mathbf{B}|}$ on $S^2$.

We are interested in the evolution of the eigenstate $| \chi \rangle $ on the Bloch sphere under a cyclic adiabatic evolution of the parameters.
The nature of this evolution will be sensitive not only to the path in the parameter space,  but also to the strengths $\left(c_1, c_2, c_3, c_4\right)$ of the various potentials in the driving protocol $\Pi$.  We shall first consider the protocol with $\left(c_1, c_2, c_3, c_4\right)= (1,1,1,1)$ to study the adiabatic evolution of eigenstates.

The boundary of the unit square  parameter space are the line segments 
$\alpha =0$, $\beta =0$, $\alpha =1$ and $\beta =1$. Further, we have the leading diagonal $\alpha = \beta$.
We observe that the direction of the magnetic field $\hat{B}$ remains invariant for the evolution of the parameters on any of these segments.
This is schematically represented in Fig.~\ref{fig:sphere}, where these five directions are shown as five points on the surface of a unit sphere. 
The existence of these subspaces for which the magnetic field has the same direction shall crucially provide us with a way to choose paths in the domain of parameters such that the synthetic magnetic field returns to its original value after traversing the path.
The table in Fig.~\ref{fig:sphere} shows the actual magnetic fields having a constant direction. The rise and fall of the magnitude of the magnetic field between the \emph{diabolical points} at the ends along these lines form an inverted parabolic shape symmetric about the center of the lines.

\begin{figure*}
\includegraphics[width=12cm]{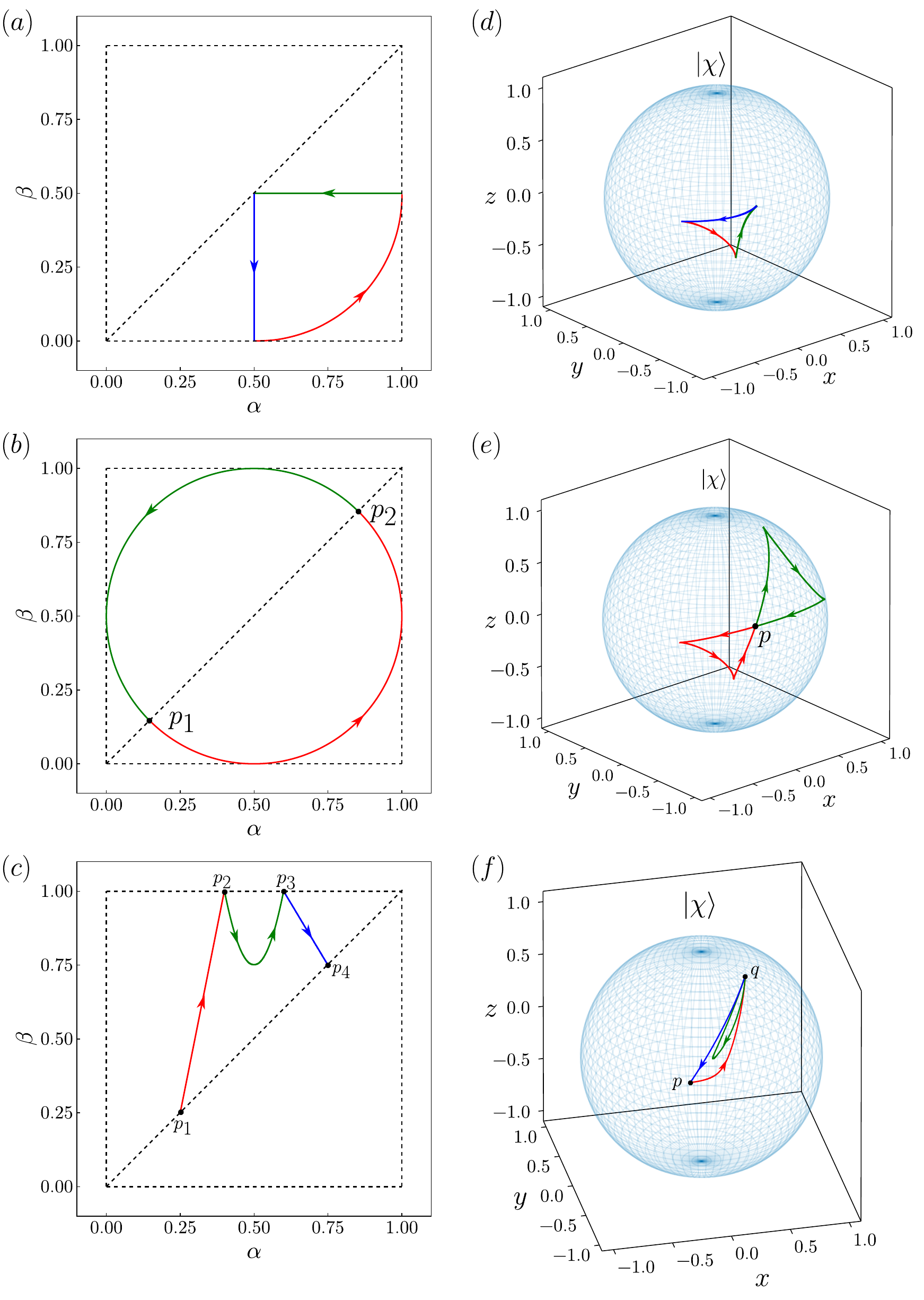}
\caption{$\left(a\right)$ A path that does not cross any of the dashed segments more than once. $\left(b\right)$ A path that crosses $\alpha = \beta$ twice at $p_1$ and $p_2$.  $\left(c\right)$ A path that crosses $\alpha = \beta$ twice at $p_1$ and $p_4$ and $\alpha = 1$ twice at $p_2$ and $p_3$.
Figures $\left(d\right)$ , $\left(e\right)$ and $\left(f\right)$ depict the corresponding evolution of the eigenstate $|\chi\rangle$ on $S^2$ for each path on the left.
}
\label{fig:path}
\end{figure*}

We investigate the possibility of having the eigenstate $|\chi \rangle$ pick up a finite geometrical phase for a cyclic adiabatic evolution of the parameters in the parameter space. 
The geometric phase picked up is given by $-\Omega/2$, where $\Omega$ is the solid angle subtended by the closed path traversed by $\hat{n}$ on the Bloch sphere.
It is, in principle, possible to coherently interfere the states after their evolution along two different paths as a way to detect this phase \cite{aharonov1959}.

Thus, any back-and-forth evolution of $(\alpha, \beta) $  on each of the four segments of the square boundary and also the leading diagonal will not lead to any phase picked up by  $|\chi \rangle$ since they correspond to fixed directions of 
$\hat{n}$. 
Further, any closed path in the parameter space that crosses any of these five segments more than once will correspond to closed paths that cross itself on $S^2$.
The crossing point on the sphere corresponds to the two points where $\hat{n}$ is the same owing to $(\alpha, \beta)$ lying on any of these five segments.
Thus, closed paths in the parameter space that cross these segments multiple times at different places will lead to one or many crossings of  $\ket{\chi}$ in $S^2$. Noting that a closed trajectory  in the parameter space is mapped to a closed path on the Bloch sphere, the number of crossings is an integer which characterizes the map from $S^1 \to S^1$ since $\pi_1\left( S^1\right) = \mathbb{Z}$. 

  Figure \ref{fig:path} shows an adiabatic evolution of $(\alpha, \beta) $ which touches the segments only once.
  The path is  parameterized as 
  \[
    \vec{R}(\tau) = \begin{cases*}
      \left(1-2 \tau, 0.5\right) , & $0 \leq \tau \leq 0.25$  \\
      \left(0.5, 1 - 2\tau \right) , & $0.25 \leq \tau \leq 0.5$  \\
      \left(\tau, 0.5 - \sqrt{\tau - \tau^2}\right) , & $0.5 \leq \tau \leq 1$
    \end{cases*} 
  \]
and shown in Fig.~\ref{fig:path}(a). The evolution of the eigenstate $|\chi \rangle $ on the Bloch sphere is shown in Fig.~\ref{fig:path}(d), which is a closed path that winds only once, owing to the fact that the five segments in the parameter space are not crossed multiple times by $\vec{R}(\tau)$. The solid angle subtended by the closed path on the sphere is the non-zero Berry's phase picked up by the eigenstate under this evolution.
Figure \ref{fig:path}(b) shows a circular path in the parameter space given by
  \[
    \vec{R}(\tau) = \left(\frac{1}{2} + \frac{1}{2}\cos(2\pi \tau), 
    \frac{1}{2} + \frac{1}{2}\sin(2\pi \tau)\right), \quad
    0 \leq \tau \leq 1.
  \]
This path crosses one of the five segments, namely the leading diagonal at the points $p_1$ and $p_2$. Since $p_1$ and $p_2$ correspond to the same eigenstate $p$ on $S^2$, the path on $S^2$ winds twice as $\vec{R}\left(\tau\right)$ winds once in the parameter space as shown in Fig.~\ref{fig:path}(e). In general, depending on the number of times a simple closed path on the parameter space intersects the dashed segments, we could have multiple crossings of the eigenstate on $S^2$. The geometric phase picked would depend on the orientation of the path on $S^2$ as it evolves. In principle, the path in the parameter space can be constructed so that there is no net geometric phase picked up under such an evolution.  

We next consider an open path $p_1 \rightarrow p_2 \rightarrow p_3 \rightarrow p_4$ on the parameter space that corresponds to an adiabatic evolution of the state with multiple windings, as shown in Fig.~\ref{fig:path}$\left(c\right)$ given by
  \[
    \vec{R}(\tau) = \begin{cases*}
    \left(\tau ,5\tau -1\right), & $0.25 \leq \tau \leq 0.4$ \\
    \left(\tau, 25\tau^2 - 25\tau +7 \right), & $0.4 \leq \tau \leq 0.6$ \\
    \left(\tau, -\frac{5}{3}\tau + 2\right), & $0.6 \leq \tau \leq 0.75$\\
    \end{cases*}.
  \]
This path symmetrically intersects two of the dashed segments at $p_1, p_4$ and $p_2, p_3$ respectively. This would result in the eigenstate winding twice on $S^2$ as seen in Fig.~\ref{fig:path}(f). Thus, even for an open path in the parameter space, we can have the eigenstate cyclically trace a closed path on $S^2$ and pick up a phase in the process. 
The points $p_1$ and $p_4$ lie on a contour of constant magnitude of the magnetic field $\abs{\mathbf{B}} = 1.24$. This implies that both the magnitude and the direction of the synthesized magnetic field return to the initial value at the end of the evolution.

Let us now consider a driving protocol with $(c_1, c_2, c_3, c_4) = (0,1,0,1)$. We have seen earlier that this yields a synthetic magnetic field $\mathbf{B} = B_o(\alpha, \beta) \hat{x}$. The magnetic field always points towards the $\hat x$  direction irrespective of $(\alpha, \beta)$, whereby the direction of the engineered magnetic field shall remain very stable to any change in these parameters. Since the magnetic field does not change its direction, a cyclic evolution of $(\alpha, \beta)$ in the parameter space shall not generate any geometric phase for the eigenstates.

\section{Discussion}
\subsection {Winding number}
We have seen that for the protocol $(c_1, c_2, c_3, c_4)=(1,1,1,1)$, a closed trajectory in the parameter space that crosses the leading diagonal (the line $\alpha = \beta$) or any of the boundary lines shall cause the eigenstate of the effective Hamiltonian to loop more than one on the Bloch sphere. The winding number for the eigenstate is then determined by the number of times the trajectory crosses these lines. We have further established in the last section that the energy cost of a trajectory is path-dependent. Since this is a smooth function of the path $\vec{R}(\tau)$, we claim that for two trajectories that are almost the same, the energy cost shall also be almost the same with a difference that shall be proportional to the tiny difference of the path lengths. We demonstrate this using the trajectory in Fig.~\ref{fig:path}(a)
\begin{figure}[htbp!]
    \includegraphics[width=0.77\linewidth]{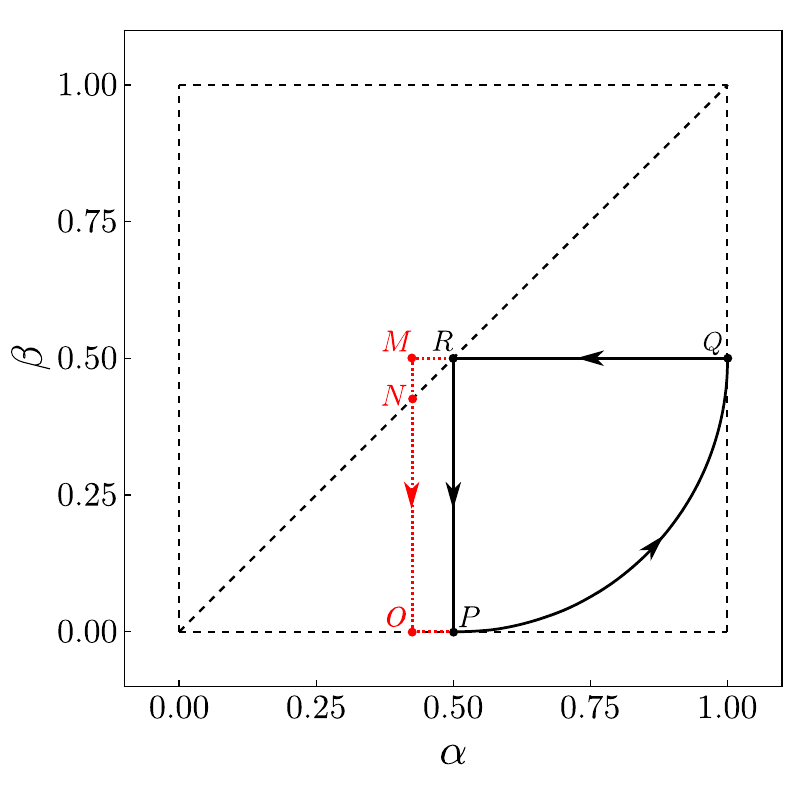}
  \caption{Two closed paths $MNOPQR$ and $PQR$ in the parameter space which differ marginally. However, for path $MNOPQR$ the winding number  is 2 and the eigenstate will cross itself twice, while for the path  $PQR$ it is 1.}
    \label{fig:winding}
\end{figure}

In Fig.~\ref{fig:winding}, the two closed paths  $MNOPQR$ and $PQR$ in the parameter space differ slightly. Thereby the energy costs for evolution along these paths will almost be the same. However, the path $MNOPQR$ crosses the diagonal line at two points, and thus, for this path, the eigenstate shall have a winding number of two. 
This demonstrates that it is possible to generate higher winding numbers by smooth variation of the path in parameter space, which does not cost much energy. An integer jump in the winding number is intrinsically related to the degeneracy structure of the Hamiltonian in the parameter space.
The system chosen by us is devoid of sufficient complexity for physical quantities to undergo a real phase transition with topological significance \cite{kolodrubetz2018, nathan2021, thouless1983}. The model is aimed simply to demonstrate the possibility of utilizing temporal partitioning as a means to synthesize a non-trivial degeneracy structure in the parameter space. The topological quantity namely the winding number just gives us the number of times the synthesized magnetic field $\vec{B}$ returns back to its original orientation while we go around only once in the parameter space could hold physical significance.

\subsection{Generalized protocols and enhanced parameter space}
\begin{figure}[htbp!]
\includegraphics[width=1.0\linewidth]{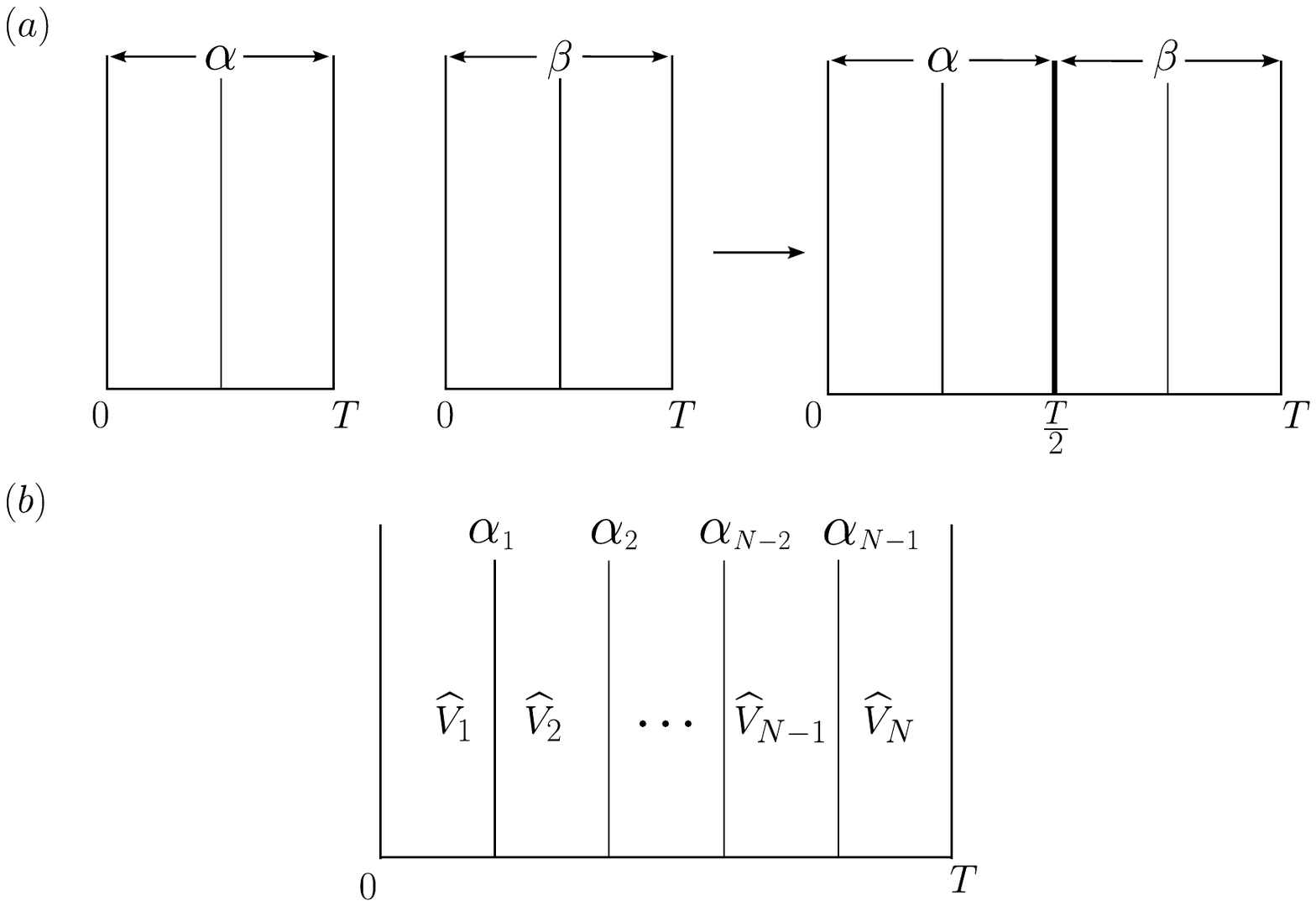}
  \caption{$\left(a\right)$ Depicts the construction of our protocol through two copies of a simpler single parameter protocol.  $\left(b\right)$ The generalized protocol with $N$ driving potentials and overlapping time spaces.}
     \label{fig:genprot}
\end{figure}

The slicing of the time period $T$ using multiple parameters is the essence of our approach. 
The simplest partitioning of a time period can be achieved using a single parameter, as shown in Fig.~\ref{fig:genprot}. The protocol that we have constructed in Eq.~(\ref{eq:protocol}) is a simple extension of this with two parameters. However, the two parameters $\alpha$ and $\beta$ vary in independent time intervals. We could include the possibility of parameters varying in intervals that overlap within the full period. 

Thus, we can generalize the protocol to have $N$ overlapping time spaces characterized by $N-1$ parameters $( \alpha_1, \alpha_2, \dots, \alpha_{_{N-1}})$ where each $ 0 \leq \alpha_n \leq 1$ such that the driving potential is 
given by:
\begin{align}\label{eq:protocolgen}
&\widehat{V}(t) = \begin{cases*}
  \widehat{V}_1 &  $f_0 T \leq t \leq f_1 T$  \\
  \widehat{V}_2 & $f_1 T \leq t \leq f_2 T$ \\
  \vdots & \vdots \\
  \widehat{V}_n & $f_{n-1} T \leq t \leq f_{n} T$\\
  \vdots& \vdots \\
  \widehat{V}_N &$f_{N-1}T \leq t \leq f_N T$
\end{cases*}, 
\end{align}
where $f_0 = 0, f_N = 1$ and $f_n$'s are given by the inclusion-exclusion principle as 
\begin{align}
    f_n = \sum_{1\leq i \leq n} \alpha_i &- \sum_{1 \leq i < j \leq n} \alpha_i \alpha_j 
     +\sum_{1 \leq i < j < k \leq n} \alpha_i \alpha_j \alpha_k  \nonumber\\
     &+~\cdots ~+ \left(-1\right)^{n+1} \alpha_1 \alpha_2 \cdots \alpha_n  
\end{align}

The coefficients of the commutators $[\widehat{V}_r, \widehat{V}_s]$ appearing in the effective Hamiltonian in Eq.~(\ref{eq:master}) are functions $\mathscr{P}_{rs}\left(\alpha_1, \dots,\alpha_{_{N-1}} \right)$ which shall contain $\left(f_n\right)^3$ as the highest order term. Thus, we could, in general, have expressions of the form $\sim \alpha^3_1 \alpha^3_2 \dots \alpha^3_{_{N-1}}$. This is different from the terms that we would get using the protocol in Eq.~(\ref{eq:protocol}), where at the highest order, we would have terms in the expansion of $\left(\alpha + \beta\right)^3$, but not $\alpha^3\beta^3$. However, for the generalized protocol Eq.~(\ref{eq:protocolgen}) with $N = 4$, we have the possibility of having terms like $\alpha^3_1\alpha^3_2\alpha^3_3$. The increased degree of the multivariate polynomials for this generalized protocol thus points towards a greater number of degeneracies in the effective Hamiltonian. 

Hence, we can expect the band structure of the effective Hamiltonian obtained from this protocol to not only be in a larger parameter space but also have an enriched structure without requiring additional driving potentials. 

\subsection{Time scales}
The problem at hand presents us with three time scales. Firstly, we have the time scale $\omega_{0}^{-1}$ set by the energy scale  of the unperturbed Hamiltonian $\widehat{\mathcal{H}}_0$. Then, we have the time period of the high frequency driving $\omega^{-1}$ and finally we have the adiabatic time scale $\Omega^{-1}$. In order for these timescales to be decoupled from each other, we would expect that $\omega \gg \omega_o \gg  \Omega$.
For a quantitative estimate of these timescales in the problem under consideration, we resort to the various energy scales in the problem.

We have already seen that $\omega_0$ is set by the matrix elements $\langle m | \widehat{\mathcal{H}}_0 | n \rangle$, where $|m\rangle, |n\rangle$ are eigenstates of  $\widehat{\mathcal{H}}_0 $. Our formalism rests upon the existence of a suitable timescale (high frequency driving) such that the original time dependent problem reduces to an effective time independent problem.
The frequency $\omega$ has to be chosen with a tolerance set by the validity of Eq.~(\ref{eq:master}). Averaging over such time scales $\left(\sim \frac{1}{\omega}\right)$ thus gives the effective averaged  contribution of the potential within a period.
\begin{equation} \Delta E_{\text{fast}} =  \frac{\omega}{2\pi}\int^{2\pi/\omega}_0 dt~\langle \psi  | \mathcal{H} (t) | \psi \rangle ,  \end{equation}
where, $| \psi \rangle  $ is some general state of the system.
The effective Hamiltonian takes into account the averaging over the fast time period. If we allow the $\widehat{\mathcal{H}}_{\text{eff}}$ to undergo slow modulation owing to slow changes in $\alpha, \beta$ then under the assumption that the two timescales are widely separated and therefore decoupled. we can, define
a slow energy scale $\Delta E_{\text{slow}}$. The slow adiabatic time scale, is set by the time it takes for the parameters to change in the parameter space along some specified path $\vec{R}(\tau) = ( \alpha(\tau), \beta(\tau) ) $ and is given by the integral of $\Braket{ \frac{d\widehat{\mathcal{H}}_{\text{eff}}\left(R(\tau)\right)}{d\tau}}$ over the adiabatic timescale. Thus  
\be 
\Delta E_{\text{slow}} = \int_0^{2\pi/\Omega} d\tau~  
\left \langle \psi  \left | {\nabla} \widehat{\mathcal{H}}_{\text{eff}}
\cdot  \frac{d \vec{R}}{d \tau} \right | \psi  \right  \rangle,
\ee
where $\nabla = ( \partial_{\alpha}, \partial_{\beta})$ is the gradient in the parameter space.

If we consider the protocol  in Eq.~(\ref{eq:protocol}), with parameters $(c_1, c_2, c_3, c_4)=(1,1,1,1)$ and  start the  time evolution from an initial point $(\alpha\left(\tau_0\right), \beta\left(\tau_0\right))$ in the parameter space,  we have $\Delta E_{\text{fast}} = \langle \psi | \mathbf{S}\cdot \mathcal{B}_f |\psi \rangle$, where
 \begin{equation}
   \mathcal{B}_f = \frac{1}{2}\begin{bmatrix}
 \alpha_0 - \beta_0 \\
 1 - 2\alpha_0\\
3 \beta_0 - \alpha_0  - 1
 \end{bmatrix}.
 \end{equation}
Here, we have assumed that during this fast evolution, the parameters do not change and remain fixed to their initial values.
For the same protocol, the energy cost for the slow evolution is given by
\begin{widetext}
\begin{equation}
 \Delta E_{\text{slow}} = \int^{\tfrac{2\pi}{\Omega} + \tau_0}_{\tau_0} d\tau \left\langle \psi \left|\mathbf{S}\cdot\mathcal{B}_1\left(\tau\right)\frac{d\alpha\left(\tau\right)}{d\tau} + \mathbf{S}\cdot\mathcal{B}_2\left(\tau\right)\frac{d\beta\left(\tau\right)}{d\tau} \right|\psi\right\rangle, 
\end{equation}
\begin{equation*}
    \text{with,}\quad \mathcal{B}_1 = -\frac{\pi}{8\omega}\begin{bmatrix}
    -12 \alpha \beta + 10 \alpha + 6 \beta ^2 -5 \\
    -4 \alpha \beta + 6 \alpha + 2\beta ^2 -3 \\
    -4\alpha \beta -2\alpha + 2 \beta ^2 + 1 \\
    \end{bmatrix}, \quad
    \mathcal{B}_2 = -\frac{\pi}{8\omega}\begin{bmatrix}
    -6\alpha^2 + 12\alpha \beta - 6\beta + 3 \\
    -2\alpha^2 + 4\alpha \beta + 6\beta - 3 \\
    -2\alpha^2 + 4\alpha \beta - 2\beta + 1 \\
    \end{bmatrix},
\end{equation*}
\end{widetext}
where the integral is evaluated along a specified path $\vec{R}(\tau) = ( \alpha(\tau), \beta(\tau))$.

It follows from the conservation of energy that the energy cost for the slow evolution $\Delta E_{\text{slow}}$ is a large multiple of the typical energy cost per fast driving cycle $\Delta E_{\text{fast}}$. Thus, the adiabatic condition on the frequencies $\Omega$ and $\omega$ can be obtained from the inequality, $\Delta E_{\text{fast}} \ll \Delta E_{\text{slow}}$.
\subsection{Summary}
In this work, we have formulated a method to adiabatically modulate driving protocols in periodically driven quantum systems.
We began with a parametrized fast-driving protocol and computed the perturbative corrections to the effective Hamiltonian and the micro-motion kick operators. Then, we induce a second, much slower drive by an adiabatic variation of these parameters. We study the effects of this variation through a simple spin $1/2$ system. We note the presence of various \emph{diabolical points} and \emph{diabolical loci} in the spectrum of this Hamiltonian. We further deduce various properties of the eigenspace of this Hamiltonian through its connection to the parameter space by constructing various adiabatic paths.
We emphasise that while adiabatic evolution in Hamiltonians can be brought about in numerous ways, our formulation of using the temporal partitions of the driving protocol as the parameters allows for arbitrarily large and complex parameter spaces without requiring additional couplings or operators. It is also important to note that the structure of degeneracy is related to the zeros of certain polynomials of these partitioning parameters and not the properties of the operators involved in the driving protocol. This has the potential to adjust energy gaps in physically relevant systems simply through tuning the parameters and/or increasing the number of parameters. While we have worked with only two parameters, we have provided a generalization of our parametrization of the driving protocol, which could potentially yield a much more enriched energy spectrum and distribution of degeneracies in the parameter space.
We observe that as a consequence of such degeneracy structures, a small deformation to the path in the parameter space could yield a new path in the Hilbert space that is not homeomorphic to the original path on the Hilbert space. Such small deformations have a negligible energy cost for the driving agency. However, they bring about discrete changes in the topological properties (crossings) of the eigenstates. In the simplest case of the spin-1/2 system, such crossings imply that for small changes in the cyclic path in the parameter space, the direction of the synthesized magnetic field may loop back to itself multiple times, which may be physically significant.
We also note that if the potentials themselves were related to, say, a lattice model or if the parameters $\alpha, \beta$ could be interpreted in the Fourier space as $k_x, k_y$ whereby the energy spectrum becomes a band structure in the Brillouin zone, then the rich structure of the diabolical points and loci in that context may actually lead to physically observable properties by choosing appropriate trajectories on the parameter space.
Finally, we comment on the regime where our analysis is valid by comparing the different energy scales involved.
We conclude by noting that the method developed in this work can be applied to any periodically driven system, whereby the topological aspects pointed out by us can find diverse ramifications in theoretical or experimental contexts.
\section{Acknowledgments}
The authors would like to thank Aritra Banerjee and Pritam K. Jana for useful discussions and suggestions. A.M. would like to thank Abhirath Anand for help with Julia. Figures were created using Makie.jl and Inkscape. 
\bibliography{bibliography.bib}

\end{document}